\documentclass{moriond}

\usepackage{amssymb,amsmath}

\def\be{\begin{equation}}
\def\ee{\end{equation}}
\def\bea{\begin{eqnarray}}
\def\eea{\end{eqnarray}}

\renewcommand{\d}{\mathrm{d}}
\newcommand{\bm}[1]{{\bf #1}}

\newcommand{\mc}[1]{\mathcal{#1}}

\renewcommand{\d}{\mathrm{d}}
\newcommand{\sT}{{\scriptscriptstyle T}}
\newcommand{\pT}{\bm{p}_\sT}
\newcommand{\kT}{\bm{k}_\sT}
\newcommand{\qT}{\bm{q}_\sT}

\renewcommand{\Re}{\text{Re}}

\newcommand{\Photo}{}

\begin{document}
\vspace*{4cm}
\title{MEASURING THE SPIN AND PARITY OF A RESONANCE IN THE $\gamma\gamma$ DECAY CHANNEL}

\author{WILCO J. DEN DUNNEN}

\address{Inst.\ for Theoretical Physics, 
Universit\"{a}t T\"{u}bingen, 
Auf der Morgenstelle 14, 
72076 T\"{u}bingen, Germany}

\maketitle

\abstracts{
We present a way to determine the spin and parity of a resonance produced 
through gluon fusion with a decay to a $\gamma\gamma$ pair based on the transverse momentum
and Collins-Soper $\phi$ distribution. 
This method also allows one to distinguish between various non-minimal coupling spin-2 scenarios
and can be used in parallel to the `standard' method based on the polar angle $\theta$.
}

\section{Introduction}

In July 2012 the ATLAS and CMS collaborations announced the discovery
of a new resonance~\cite{Aad:2012tfa,Chatrchyan:2012ufa} in their search for the Standard Model (SM) Higgs boson.
Decays of this resonance to $ZZ$~\cite{Aad:2013xqa,Chatrchyan:2013mxa} and
$\gamma\gamma$~\cite{Aad:2013xqa} have been established at over 5 sigma,
whereas strong evidence ($\sim 4\sigma$) exists for a decay to $WW$~\cite{Aad:2013xqa,Chatrchyan:2013iaa}.
Besides these channels, first measurements of the decay to $\tau\tau$ ($3 \sigma$)~\cite{Chatrchyan:2014nva}
and to $bb$ ($2 \sigma$)~\cite{Chatrchyan:2013zna} have now also been published.

Measurements of the spin of the resonance exclude a minimal coupling spin-2 resonance 
produced through gluon fusion in the $ZZ$ channel at approximately $2\sigma$~\cite{Chatrchyan:2013mxa}.
The same scenario is excluded at almost $3\sigma$ in the $\gamma\gamma$ 
channel~\cite{Aad:2013xqa} and at $2\sigma$ in the $WW$ channel~\cite{Aad:2013xqa}. 
Exclusions in the remaining channels or of non-minimal coupling spin-2 scenarios have not yet been presented.

Regarding the parity of the resonance, the option of a pseudoscalar in the $ZZ$ channel
has been excluded at approximately $2.5\sigma$~\cite{Aad:2013xqa}, 
whereas in the $WW$ channel this can only be done at $1\sigma$~\cite{Chatrchyan:2013iaa}.
In the $\gamma\gamma$ channel no parity determination can be made using conventional 
methods~\cite{Choi:2002jk,Gao:2010qx,Bolognesi:2012mm,Choi:2012yg,Ellis:2012jv}
as the hard scattering $gg\to h\to\gamma\gamma$, which is characterized by
only one single angle $\theta$, is independent of the parity of $h$.

We will discuss here a method to determine both the parity of a resonance 
in the $\gamma\gamma$ channel~\cite{Boer:2011kf} and to distinguish between various spin-2 coupling
scenarios that could not be distinguished on the basis of the $\theta$ distribution alone~\cite{Boer:2013fca}.
As we will show, various spin-2 scenarios can be distinguished on the basis of the Collins-Soper 
$\phi$ distribution, whereas the parity of the resonance manifests itself through the transverse momentum distribution.
The effects on the transverse momentum distribution are small, but the effect on the $\phi$ distribution,
for various spin-2 scenarios, is large enough to be measurable with the currently recorded data set.

The underlying principle of these methods relies on the fact that gluons are
linearly polarized in the direction of their transverse momentum when entering the hard scattering.
This polarization can be generated perturbatively, but it can also have a non-perturbative (intrinsic) component. 
It was realized that the perturbatively generated polarization forces one to modify the standard 
(quark initiated) Drell-Yan $q_\sT$-resummation procedure~\cite{Nadolsky:2007ba,Catani:2010pd}
and its effects on SM Higgs boson production have since been taken into account~\cite{Catani:2011kr,deFlorian:2011xf,deFlorian:2012mx,Catani:2013tia}.
We will employ Transverse Momentum Dependent (TMD) factorization
to systematically take into account both perturbative \emph{and} non-perturbative gluon polarization
and calculate the effect on arbitrary colorless spin-0 and spin-2 boson production.

\section{The $pp\to X_{0,2}X \to\gamma\gamma X$ differential cross section in TMD factorization}

Within TMD factorization the full $pp\to X_{0,2}X \to\gamma\gamma X$ cross section,
for a gluon fusion initiated process, is split into a partonic
$gg\to \gamma\gamma$ cross section and two TMD gluon correlators that describe the 
distribution of gluons inside the proton as a function of their longitudinal and transverse momentum.
More specifically, the differential cross section is written as 
\cite{Ji:2005nu,Sun:2011iw,Ma:2012hh},
\begin{equation}\label{eq:factformula}
\frac{\d\sigma}{\d^4 q \d \Omega}
  \propto 
  \int\!\! \d^{2}\pT \d^{2}\kT
  \delta^{2}(\pT + \kT - \qT)
  \mc{M}_{\mu\rho}
  \left(\mc{M}_{\nu\sigma}\right)^*
  \\
  \Phi_g^{\mu\nu}(x_1,\pT,\zeta_1,\mu)\,
  \Phi_g^{\rho\sigma}(x_2,\kT,\zeta_2,\mu),
\end{equation}
with the longitudinal momentum fractions $x_1={q\cdot P_2}/{P_1\cdot P_2}$
and $x_2={q\cdot P_1}/{P_1\cdot P_2}$, $q$ the momentum of the photon pair, 
$\mc{M}$ the $gg\to \gamma\gamma$ partonic hard scattering matrix element
and $\Phi$ the following gluon TMD correlator in an unpolarized proton,
\begin{align}\label{eq:TMDcorrelator}
\Phi_g^{\mu\nu}(x,\pT,\zeta,\mu) 
&\equiv
	      \int \frac{\d(\xi\cdot P)\, \d^2 \xi_\sT}{(x P\cdot n)^2 (2\pi)^3}
	      e^{i ( xP + p_\sT) \cdot \xi}
	      \left\langle P \left| F_a^{n\nu}(0)
	      \left(\mc{U}_{[0,\xi]}^{n[\text{--}]}\right)_{ab} F_b^{n\mu}(\xi)
	      \right|P \right\rangle \Big|_{\xi \cdot P^\prime = 0}\nonumber\\
&=	-\frac{1}{2x} \bigg \{g_\sT^{\mu\nu} f_1^g(x,\pT^2,\zeta,\mu)
	-\bigg(\frac{p_\sT^\mu p_\sT^\nu}{M_p^2}\,
	{+}\,g_\sT^{\mu\nu}\frac{\pT^2}{2M_p^2}\bigg)
	h_1^{\perp\,g}(x,\pT^2,\zeta,\mu) \bigg \},
\end{align}
with $p_{\sT}^2 = -\pT^2$ and $g^{\mu\nu}_{\sT} = g^{\mu\nu}
- P^{\mu}P^{\prime\nu}/P{\cdot}P^\prime - P^{\prime\mu}P^{\nu}/P{\cdot}P^\prime$,
where $P$ and $P^\prime$ are the momenta of the colliding protons and $M_p$ their mass.
The gauge link $\mc{U}_{[0,\xi]}^{n[\text{--}]}$ for this process arises 
from initial state interactions. 
It runs from $0$ to $\xi$ via minus infinity along the direction $n$, 
which is a time-like dimensionless four-vector with no transverse 
components such that $\zeta^2 = (2n{\cdot}P)^2/n^2$. 
With the appropriate choice of $\zeta$ and $\mu$, the usual soft factors in Eqs.\
\eqref{eq:factformula} and \eqref{eq:TMDcorrelator} are absorbed into the
TMD correlators~\cite{Ji:2005nu,Ma:2012hh} and the hard part is free of large logs.
The second line of Eq.\ \eqref{eq:TMDcorrelator} contains the parametrization~\cite{Mulders:2000sh}
of the leading twist contributions to the TMD correlator,
where $f_1^g$ is the unpolarized gluon distribution and
$h_1^{\perp\,g}$ the linearly polarized gluon distribution.

The general structure of the differential cross section follows from Eq.\ \eqref{eq:factformula} 
and \eqref{eq:TMDcorrelator} and can be written as~\cite{Qiu:2011ai}
\begin{multline}\label{eq:genstruc}
\frac{\d\sigma}{\d Q \d Y \d^2 \qT\, \d\cos\theta\, \d\phi} 
  \propto 
	F_1\, \mc{C} \left[f_1^gf_1^g\right]
	+ F_2\,	\mc{C} \left[w_2\, h_1^{\perp g}h_1^{\perp g}\right]
	+ F_3\, 	\mc{C} \left[w_3 f_1^g h_1^{\perp g}
				  + (x_1 \leftrightarrow x_2) \right]\cos(2\phi)\\
	+ F_3^{\prime}\, \mc{C} \left[w_3 f_1^g h_1^{\perp g} 
				  - (x_1 \leftrightarrow x_2) \right]\sin(2\phi)
	+ F_4\,	\mc{C} \left[w_4\, h_1^{\perp g}h_1^{\perp g}\right]\cos(4\phi),
\end{multline}
up to corrections that are $\qT^2/Q^2$ suppressed at small $\qT$.
The cross section is differential in $Q$, $Y$ and $\qT$, which are the invariant mass,
rapidity and transverse momentum of the pair in the lab frame
and in the Collins-Soper angles $\theta$ and $\phi$.
The latter two are defined as the polar and azimuthal angle
in the Collins-Soper frame \cite{Collins:1977iv},
which is the diphoton rest frame with the $\hat{x}\hat{z}$-plane spanned by the 3-momenta 
of the colliding protons and the $\hat{x}$-axis set by their bisector. 
The convolution $\mc{C}$ is defined as
\begin{equation}
\mathcal{C}[w\, f\, g] \equiv \int\! \d^{2}\pT\int\! \d^{2}\kT\,
  \delta^{2}(\pT+\kT-\bm q_{\sT})
  w(\pT,\kT)\, f(x_{1},\pT^{2})\, g(x_{2},\kT^{2}),
\end{equation}
in which the longitudinal momentum fractions are given in 
the aforementioned kinematical variables by
$x_{1,2} = e^{\pm Y} \sqrt{(Q^2 + \qT^2)/s}$.
The weights in the convolutions are defined as
\begin{align}
 w_2		&\equiv \frac{2 (\kT{\cdot}\pT)^2 - \kT^2 \pT^2 }{4 M_p^4}, \qquad
 w_3 		\equiv \frac{\qT^2\kT^2 - 2 (\qT{\cdot}\kT)^2}{2 M_p^2 \qT^2},\nonumber\\
 w_4		&\equiv  2\left[\frac{\pT{\cdot}\kT}{2M_p^2} - 
		\frac{(\pT{\cdot}\qT) (\kT{\cdot}\qT)}{M_p^2\qT^2}\right]^2 -\frac{\pT^2\kT^2 }{4 M_p^4}.
\end{align}

Using the following parametrization of the $X_{0,2}\gamma\gamma$ interaction vertex,
\begin{align}
V[X_0\to V^\mu (q_1) V^\nu (q_2)] 		&= a_1 q^2 g^{\mu\nu} + a_3 \epsilon^{q_1 q_2 \mu\nu},\nonumber\\
V[X_2^{\alpha\beta}\to V^\mu (q_1)V^\nu (q_2)] 	&= \frac{1}{2}c_1 q^2 g^{\mu\alpha} g^{\nu\beta} 
      + \left(c_2 q^2 g^{\mu\nu} + c_5 \epsilon^{q_1 q_2 \mu\nu} \right) 
      \frac{\tilde{q}^\alpha \tilde{q}^\beta}{q^2},
 \end{align}
where $q\equiv q_1 + q_2$ and $\tilde{q}\equiv q_1 - q_2$, one finds
for a spin-0 boson up to a constant factor
\begin{equation}\label{eq:Fspin0}
 F_1 = (4|a_1|^2 + |a_3|^2)^2, \qquad
 F_2 = (4|a_1|^2 + |a_3|^2)(4|a_1|^2 - |a_3|^2),
\end{equation}
and for a spin-2 boson
\begin{align}\label{eq:Fspin2}
 F_1 &= 18 A^+ |c_1|^2 \sin^4\theta +  {A^+}^2\! \left( 1 - 6\cos^2\theta + 9\cos^4\theta \right)
	+ 9 |c_1|^4 \left(1 + 6 \cos^2\theta + \cos^4\theta \right),\nonumber\\
 F_2 &= 9\, A^- |c_1|^2 \sin^4\theta +  A^-A^+\left( 1 - 6\cos^2\theta + 9\cos^4\theta \right),\nonumber\\
 F_3 &= 6\, B^-\, \left[ A^+( 3 \cos^2\theta - 1) + 3 |c_1|^2 (\cos^2\theta + 1) \right] \sin^2\theta,\nonumber\\
 F_3^\prime &= 12\, \Re(c_1 c_5^*)\, 
		\left[ A^+( 3 \cos^2\theta - 1) + 3 |c_1|^2 (\cos^2\theta + 1) \right] \sin^2\theta,\nonumber\\
 F_4 &= 18\, |c_1|^2\, \left[B^+ + 2 |c_5|^2\right] \sin^4\theta,
\end{align}
where we have defined 
$A^\pm\equiv |c_1+4c_2|^2\pm 4|c_5|^2$, $B^\pm\equiv |c_1 + 2 c_2|^2\pm 4|c_2|^2$.

\section{Numerical results}

To make numerical predictions $h_1^{\perp g}$ will be expressed as 
$h_1^{\perp g} = \mc{P}\, 2M_p^2/\pT^2 f_1^g$,
where the degree of polarization $\mc{P}$ will be calculated
as described in an earlier publication~\cite{Dunnen:2013zua}.
For $f_1^g$ we use the same Ansatz as described before~\cite{Boer:2013fca,Dunnen:2013fxa}.
Plots are made for the benchmark scenarios commonly used in the literature~\cite{Bolognesi:2012mm},
to which we add $2_{h^\prime}^+$, $2_{h^{\prime\prime}}^+$
and $2_{\text{CPV}}$. The scenarios are summarized in Table \ref{tab:bmscenarios}.

\begin{table}[htb]
\caption{Various spin, parity and coupling scenarios.}
\label{tab:bmscenarios}
\vspace{0.4cm}
\centering
 \begin{tabular}{cccrrrrrr}
  	&$0^+$ &$0^-$ &$2_m^+$ &$2_h^+$ &$2_{h^\prime}^+$ &$2_{h^{\prime\prime}}^+$ &$2_{h}^-$ &$2_{\text{CPV}}$\\
 \hline
 \hline
 $a_1$		&1 &0 &- &- &- &- &- &-\\
 $a_3$		&0 &1 &- &- &- &- &- &-\\
 $c_1$		&- &- &1 &0 &1 &1 &0 &1\\
 $c_2$		&- &- &$-\frac{1}{4}$ &1 &1 &$-\frac{3}{2}$ &0 &0 \\
 $c_5$ 		&- &- &0 &0 &0 &0 &1 &5\\
 \end{tabular} 

\end{table} 

\begin{figure}[htb]
\begin{minipage}{0.33\linewidth}
\centerline{\includegraphics[width=\linewidth]{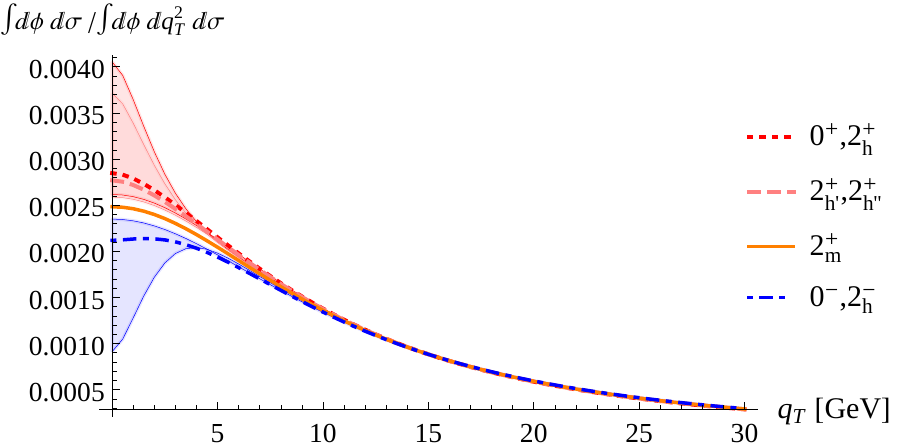}}
\end{minipage}
\hfill
\begin{minipage}{0.32\linewidth}
\centerline{\includegraphics[width=\linewidth]{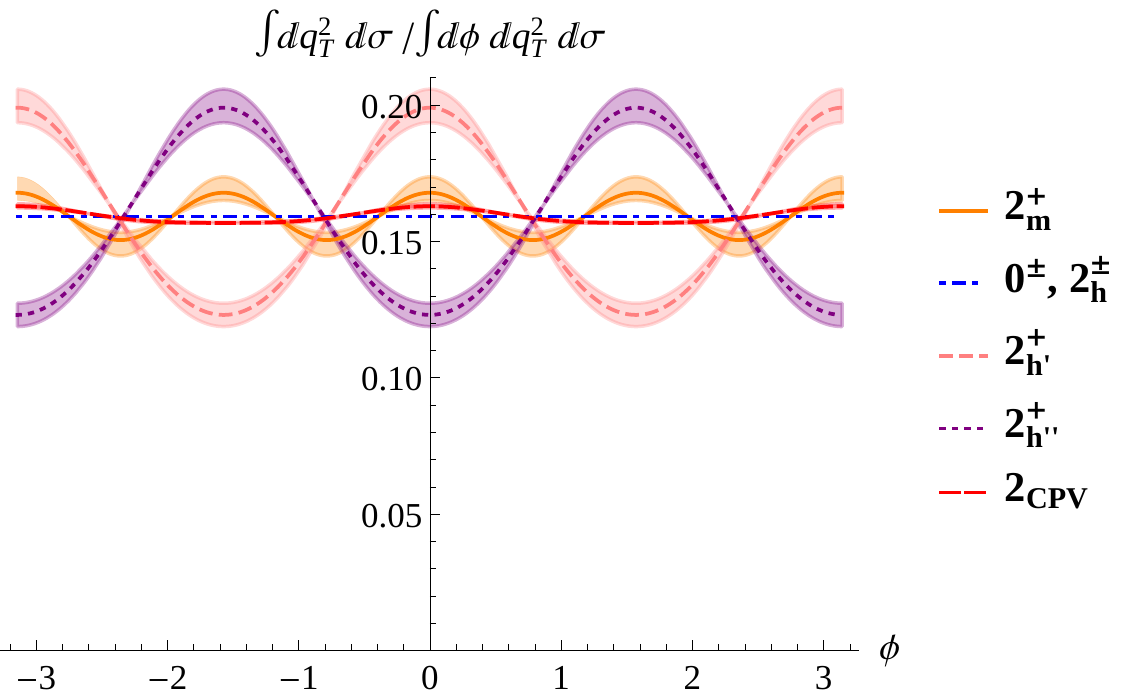}}
\end{minipage}
\hfill
\begin{minipage}{0.32\linewidth}
\centerline{\includegraphics[width=\linewidth]{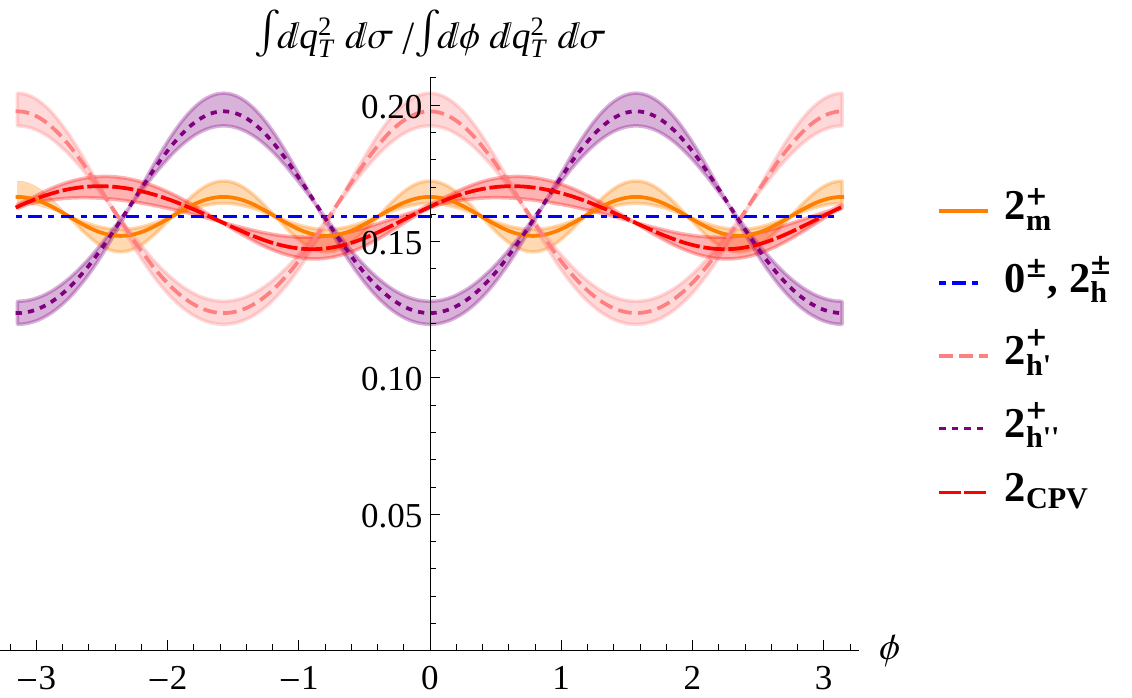}}
\end{minipage}
\caption[]{Plot of the $q_\sT^2$ distribution at $Y=0$ (left), the $\phi$ distribution at $Y=0$ (center),
and the $\phi$ distribution at $Y=1$ (right), all at $\theta=\pi/2$ for a 125 GeV resonance 
at a center of mass energy of 8 TeV.
The shaded area is due to the uncertainty in the degree of polarization.}
\label{fig:distrs}
\end{figure}

In Figure \ref{fig:distrs} we show our predictions for the $q_\sT^2$ and CS $\phi$ distributions.
Even parity states have an enhanced cross section at small $q_\sT$ with respect to negative
parity states, but the difference is small with a large uncertainty.
Including evolution of the distributions we come to the same conclusion~\cite{Boer:2014tka}.
To lower the uncertainty, a measurement of the TMDs could be made in a different process, e.g.,
$C$ even quarkonium production~\cite{Boer:2012bt} 
or $C$ odd quarkonium production in association with a photon~\cite{Dunnen:2014eta}.

The effects are larger on the CS $\phi$ distribution: 
various spin-2 coupling scenarios produce non-isotropic $\phi$ distributions
with a modulation of up to 25\%.
The $2_{\text{CPV}}$ benchmark scenario displays a characteristic
\emph{asymmetric} $\phi$ distribution in the forward region that 
can \emph{only} be caused by a $CP$-violating coupling.

\section{Conclusions}

We have calculated the $q_\sT^2$ and CS angle $\phi$ distribution in the process 
$pp\to X_{0,2} X \to \gamma\gamma X$ using TMD factorization.
The $q_\sT^2$ distribution depends on the parity of the resonance, 
but numerical predictions show that the difference is relatively small with a large uncertainty.
The CS $\phi$ distribution, on the other hand, shows large modulations, up to 25\%,
for various spin-2 scenarios, making this a realistic observable to discrimante
between spin-0 and various spin-2 scanarios.
\emph{This work was supported by the German Bundesministerium f\"{u}r Bildung und Forschung (BMBF),
grant no. 05P12VTCTG.}

\end{document}